# From Wavefunction Collapse to Superconductivity: Evolution of the Electronic State in Compressed GaNb$_4$Se$_8$

Yuejian Wang[1,*], Zhongyan Wu[2], K C Bhupendra[3], Dongzhou Zhang[4], Lin Wang[2], Sanjay V. Khare[3], Lilian Prodan[5,6], Vladimir Tsurkan[5,6]

[1]Department of Physics, Oakland University, Rochester, MI 48309, USA
[2]Center for High-Pressure Science (CHiPS), State Key Laboratory of Metastable Materials Science and Technology, Yanshan University, Qinhuangdao, Hebei 066004, China
[3]Department of Physics and Astronomy, and Wright Center for Photovoltaics Innovation and Commercialization (PVIC), University of Toledo, Toledo, Ohio 43606, USA
[4]GeoSoilEnviroCARS, University of Chicago, Argonne, IL 60439, USA
[5]Experimental Physics V, Center for Electronic Correlations and Magnetism, University of Augsburg, Augsburg 86135, Germany
[6]Institute of Applied Physics, Moldova State University, MD-2028 Chisinau, Republic of Moldova

## Abstract

Understanding how electronic transport evolves from localized to itinerant regimes in correlated cluster solids remains an important challenge in condensed-matter physics. Here we investigate the pressure-dependent transport properties of the lacunar spinel GaNb$_4$Se$_8$, a cluster Mott insulator at ambient conditions. At low pressures, the resistivity follows Efros–Shklovskii variable-range hopping, indicating Coulomb-gap-controlled carrier localization ($\xi \approx 6.1$Å). A crossover toward metallic transport begins near ~5 GPa, whereas a crystallographic transition from the cubic phase to a monoclinic C2 phase occurs at significantly higher pressure (~20 GPa), establishing a hierarchy characterized by the decoupling of electronic delocalization from structural symmetry change. At higher pressures, superconductivity ($\xi(0) \approx 80$-$90$Å) emerges from the correlated metallic regime. These results identify GaNb$_4$Se$_8$ as a platform for studying correlation-controlled transport evolution in cluster-based solids.

*Email: Ywang235@oakland.edu

*Introduction*—The Mott metal-insulator transition (MIT) represents a fundamental paradigm in condensed matter physics, where the competition between on-site Coulomb repulsion ($U$) and electronic bandwidth ($W$) governs the emergence of exotic quantum phases [1, 2]. In many correlated electron systems, superconductivity emerges from the collapse of a Mott insulating state through external tuning parameters such as chemical doping or physical pressure [3]. While the macroscopic evolution of these phases is well-documented, the microscopic structural mechanisms—specifically how local orbital degrees of freedom drive the electronic delocalization required for superconductivity—remain a subject of intense investigation [4].

Lacunar spinels with the general formula $AM_4X_8$ ($A=$ Ga, Ge; $M=$ V, Nb, Ta, Mo; $X=$ S, Se) offer a unique platform to explore these dynamics. Their crystal structure consists of molecular-like $M_4$ clusters, where electronic states are inherently localized within tetrahedral units [5-11]. The profound coupling between lattice and electronic degrees of freedom in these materials is often mediated by the Jahn-Teller (JT) effect, which lowers the local symmetry to stabilize the cluster Mott insulating state [5, 6, 12-16]. Among this family, GaNb$_4$Se$_8$ is of particular interest due to its narrow-gap insulating state and high sensitivity to compression [17]. Despite prior reports of high-pressure behavior of this compound [18, 19], the microscopic mechanism underlying the transition from a localized to a superconducting state, as well as the interplay between lattice symmetry breaking and electronic reconstruction, remains elusive.

In this letter, we report the emergence of bulk superconductivity in single-crystalline GaNb$_4$Se$_8$ and identify the underlying high-pressure structural framework. High-pressure synchrotron X-ray diffraction reveals a cubic-to-monoclinic structural transition near 20 GPa. While significant peak broadening and sample pulverization under extreme compression pose challenges for definitive refinement, a monoclinic $C2$ model provides a consistent identification of this previously unknown phase. Electrical transport measurements demonstrate a pressure-driven crossover from Efros-Shklovskii variable-range hopping to a correlated metallic regime, culminating in a zero-resistance superconducting state above 30 GPa ($Tc>5$ K at 48.3 GPa). Our analysis reveals that the pressure-induced suppression of JT distortions acts as an "orbital gate", unlocking the localized Nb $4d$ states to facilitate electronic delocalization and bandwidth broadening. These results link the restoration of local symmetry to the Mott insulator-to-superconductor transition and identify the structural precursor to emergent quantum phases in GaNb$_4$Se$_8$.

*Results and discussion*—We first investigated the structural evolution of GaNb$_4$Se$_8$ (Supplementary Note 1 and Supplementary Fig. S1 for the synthesis and characterization) using synchrotron X-ray diffraction up to 38.5 GPa (Fig. 1a) (Supplementary Note 2). At ambient pressure, sharp reflections can be indexed to a face-centered cubic (*fcc*) structure (space group $F\bar{4}3m$, $a = 10.42$ Å), a characteristic of lacunar spinels [8, 14, 18, 19]. Upon compression, the *fcc* symmetry is maintained up to ~20 GPa (Fig. 1a, c).

A structural transition onset was observed near 20 GPa, marked by the appearance of new reflections and the gradual weakening of the parent *fcc* peaks. Above 31 GPa, the transformation to a high-pressure phase is complete. Notably, the recovered sample returns to the *fcc* phase upon decompression (Fig. 1a, top), demonstrating the reversibility of this structural reconfiguration. While single-crystal X-ray diffraction (XRD) was initially sought to identify the high-pressure phase, sample brittleness led to pressure-induced pulverization (Supplementary Note 3 and Supplementary Figs. S2-4). Consequently, several candidate structural models [9, 10, 20] were evaluated against our XRD data and energy calculations. While the orthorhombic *Imm*2

structure—common in V-based analogues—is energetically unfavorable (Fig. 1b), a monoclinic $C2$ model provides an excellent fit (Fig. 1c), echoing the behavior of the isostructural $GaTa_4Se_8$ [20].

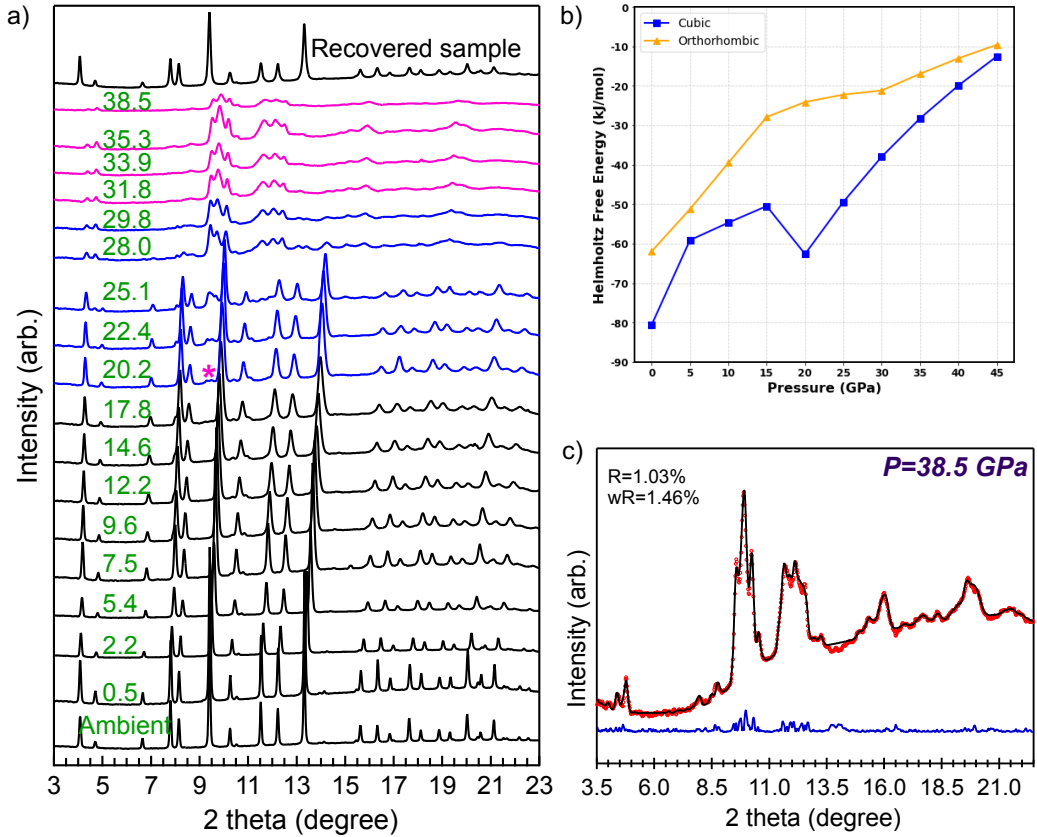

**Fig. 1.** Structural evolution of polycrystalline $GaNb_4Se_8$ under pressure. a) Synchrotron X-ray diffraction patterns collected upon compression from ambient conditions to 38.5 GPa and subsequent decompression (recovered sample). Pressures (in GPa) are indicated on the left. b) Pressure dependence of calculated Helmholtz free energies for the cubic and orthorhombic structures. c) Representative Rietveld refinements of diffraction patterns collected at 38.5 GPa. Black circles denote experimental data, red lines represent calculated profiles, and blue curves show the difference.

Following the identification of the high-pressure phase, we investigated the electronic transport properties of $GaNb_4Se_8$ under varied pressures and temperatures (Supplementary Note 4). As shown in Fig. 2a, the resistance increases exponentially upon cooling at ambient pressure, a hallmark of carrier localization within the $Nb_4$ clusters driven by strong electron–electron correlations. While the transport is often phenomenologically described by the Arrhenius activation law—yielding an activation energy of ~146.7 meV in alignment with prior reports [16, 19]— a more rigorous analysis reveals that the system is governed by the Efros-Shklovskii variable-range hopping (ES-VRH) mechanism, $\rho = \rho_0 e^{(\frac{T_0}{T})^{1/2}}$ [21] (Supplementary Note 5).

This model provides quantitative evidence for a fundamental transition in the localization state. In the high-temperature regime (220–300 K), the characteristic temperature $T_0 \approx 3016$ K corresponds to a localization length $\xi \approx 15.5$ Å and a hopping energy of $W_{ES} \approx 38.2$ meV (at 260 K), suggesting that electronic wavefunctions remain extended across multiple $Nb_4$ clusters. However, upon cooling below 100 K, the hopping energy decreases to 32.7 meV (at 75 K) as carriers optimize for lower-energy paths, but $T_0$ increases significantly to $\approx 7685$ K. Consequently, $\xi$ contracts sharply to ~6.1 Å, aligning strikingly with the inter-cluster distance ($d_{inter} \approx 6.7$ Å). This represents direct transport evidence for a 'wavefunction collapse' where carriers become strictly confined within individual $Nb_4$ tetrahedra, confirming $GaNb_4Se_8$ as a robust cluster Mott insulator.

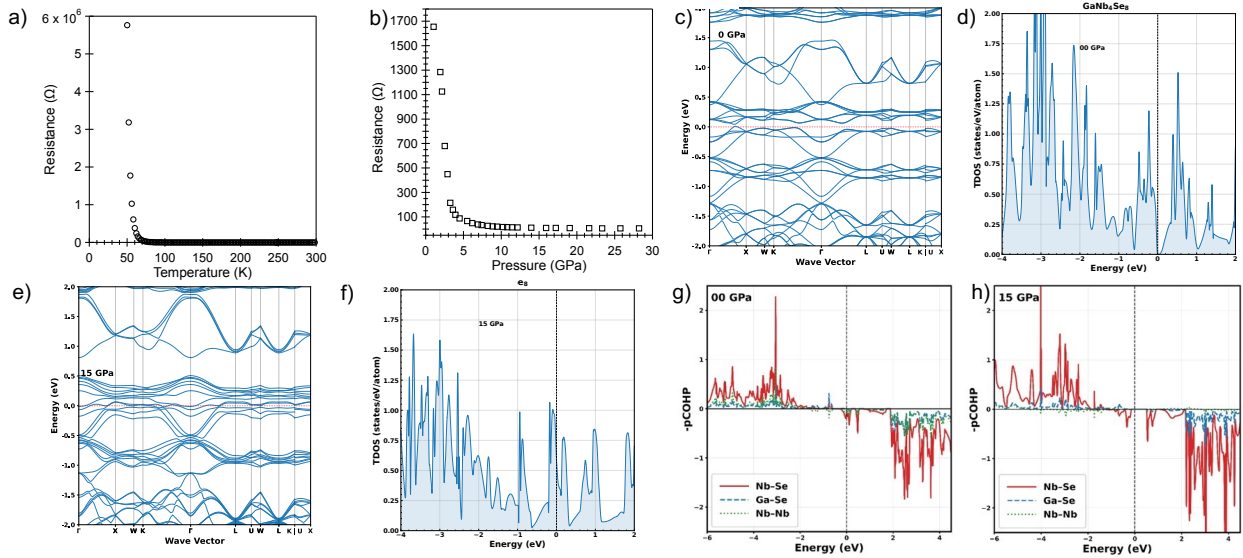

**Fig. 2.** Pressure-driven electronic evolution in $GaNb_4Se_8$. (a) Temperature-dependent resistance at ambient pressure. (b) Pressure dependence of resistance at room temperature. (c,d) Electronic band structure and density of states (DOS) of cubic $GaNb_4Se_8$ at 0 GPa. (e,f) Electronic band structure and DOS at 15 GPa. (g,h) Projected crystal orbital Hamilton population (–pCOHP) for selected Nb–Se, Ga–Se, and Nb–Nb nearest-neighbor interactions at 0 GPa and 15 GPa. Positive and negative –pCOHP values correspond to bonding and antibonding contributions, respectively. The Fermi level is set to 0 eV.

The pressure-dependent resistance at room temperature (Fig. 2b) reveals a rapid decline below ~5 GPa, reflecting the enhanced inter-cluster electronic overlap. Between 10 and 14 GPa, the temperature dependence of resistance weakens and eventually transitions to a metallic temperature coefficient, marking the crossover from a Mott insulator to a correlated metallic regime. Crucially, this electronic transition occurs well below the cubic-to-monoclinic structural transition (~20 GPa) observed via X-ray diffraction. This electronic-structural decoupling implies that the pressure-induced metallization originates primarily from bandwidth enhancement within the cubic framework, rather than being triggered by a structural symmetry breaking.

To understand the microscopic origin of the observed transport behavior and the pressure-induced metallization, density-functional calculations were performed (Supplementary Note 6). At ambient pressure, the calculated band structure (Fig. 2c) and total density of states (TDOS) (Fig. 2d) reveal a narrow energy gap separating the valence and conduction bands, aligning with the observed the cluster Mott characteristics of $GaNb_4Se_8$. Crucially, the states near the Fermi level ($E_F$) are dominated by $Nb_4$ cluster molecular orbitals, providing a theoretical basis for the carrier localization observed in transport. The electronic configuration obtained from DFT calculations provides a microscopic foundation for the 'wavefunction collapse' inferred from our ES-VRH analysis. While the experimental transport captures the temperature-dependent evolution of carrier localization, the computational ground state (0 K) reveals that the states near $E_F$ are inherently confined within individual $Nb_4$ clusters. This spatial confinement, enforced by strong electron correlations, represents the physical limit toward which the electronic wavefunctions contract as thermal fluctuations are suppressed at low temperatures.

Under compression, the bands broaden significantly and cross the Fermi level at ∼15 GPa (Fig. 2e), accompanied by a finite DOS at $E_F$ (Fig. 2f). Notably, these calculations were performed within the cubic symmetry, confirming that metallization is driven by bandwidth enhancement ($W$) sufficient to overcome the on-site Coulomb repulsion ($U$) — a finding that corroborates the electronic-structural decoupling observed experimentally.

To further elucidate this evolution, we performed projected crystal orbital Hamilton population (pCOHP) calculations (Figs. 2g–h). At ambient pressure, the average integrated COHP for Nb–Se bonds is −0.88 eV per bond, whereas Ga–Se bonds contribute only −0.013 eV per bond, confirming that the electronic states near $E_F$ are primarily governed by Nb–Se bonding interactions, while the contribution from Ga–Se bonds is negligible.

Under compression, distinct antibonding features emerge below the Fermi level near −0.5 eV (Fig. 2h), accompanied by a progressive reduction in the Nb–Se bonding strength. Quantitatively, the average integrated COHP (ICOHP) for Nb–Se bonds decrease in magnitude from −0.88 eV per bond at ambient pressure to −0.72 eV at 15 GPa, representing a reduction of approximately 15%. Consistently, the total integrated COHP summed over all bonding interactions becomes less negative with increasing pressure, evolving from −0.35 eV at 0 GPa to −0.21 eV at 15 GPa. This pressure-induced weakening of the Nb–Se bonding strength reflects a transition from rigid, covalent-like carrier confinement within the $Nb_4$ clusters toward a more delocalized regime. By facilitating enhanced inter-cluster electronic hopping, this redistribution of bonding states drives the broadening of the Nb 4$d$ bandwidth ($W$), ultimately triggering the collapse of the Mott insulating state and the emergence of the correlated metallic phase observed in our transport measurements.

To identify the structural origin of this pressure-induced delocalization, we investigated the evolution of the local coordination environment, specifically the $NbSe_6$ octahedra that constitute the Nb₄ clusters (Figs. 3a–d). At ambient pressure, the Nb atoms are significantly displaced from the centrosymmetric position of the $NbSe_6$ octahedron, as evidenced by the inequivalent Nb(1)-Se(1) and Nb(1)-Se(2) bond lengths (Fig. 3e). This structural asymmetry is a manifestation of a JT distortion, which serves to split the Nb 4$d$ orbitals and stabilize the localized Mott insulating state. Our analysis reveals a distinct two-regime evolution of these bond lengths under compression. Below ∼5 GPa, the Nb(1)-Se(2) bond contracts linearly while the Nb(1)-Se(1) bond remains nearly constant, indicating a progressive shift of the Nb atom within the octahedron. However, above ∼5 GPa, this trend reverses: the Nb(1)-Se(2) bond becomes pressure-independent, while the Nb(1)-Se(1) bond undergoes significant compression.

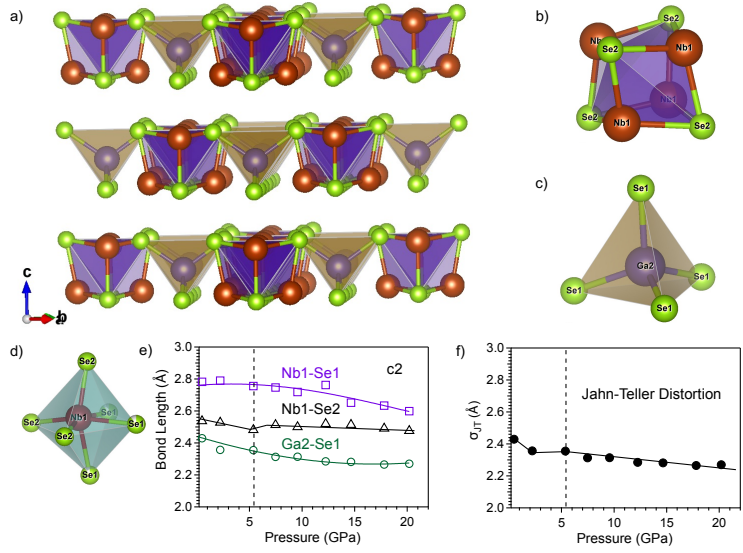

**Fig. 3** Pressure-driven modification of local bonding in GaNb$_4$Se$_8$. (a) Crystal structure of GaNb$_4$Se$_8$, composed of Nb$_4$Se$_4$ clusters and GaSe$_4$ tetrahedra. (b,c) Local structural units highlighting the Nb$_4$Se$_4$ cluster and GaSe$_4$ tetrahedron, respectively. (d) Schematic representation of the NbSe$_6$ octahedral coordination environment. (e) Pressure dependence of selected Nb–Se and Ga–Se bond lengths obtained from Rietveld refinement of synchrotron X-ray diffraction data. (f) Evolution of the JT distortion parameter as a function of pressure.

This crossover is quantitatively captured by the JT distortion parameter, $\sigma_{JT}$ defined as [22-24]

$$\sigma_{JT} = \left\{ \sum_{i=1}^{2} (L_{\text{Nb(1)-Se}(i)} - \bar{L}_{\text{Nb-Se}})^2 \right\}^{1/2}$$

Where $L_{\text{Nb(1)-Se}(i)}$ denotes the specific bond lengths between Nb and Se atoms, and $\bar{L}_{\text{Nb-Se}}$ represents the average Nb-Se bond length within the NbSe$_6$ octahedron. While $\sigma_{JT}$ remains nearly constant in the low-pressure regime, it exhibits a progressive decrease above 5 GPa, signaling the suppression of the structural distortion and a move toward a more centrosymmetric Nb coordination (Fig. 3f). Crucially, this structural crossover at ~5 GPa coincides precisely with the onset of the insulator-to-metal transition and the rapid decline in resistance observed in our transport measurements. The pressure-induced suppression of the JT distortion provides the structural basis for the redistribution of bonding states revealed by our pCOHP analysis. By restoring local symmetry, compression effectively reduces the orbital splitting and facilitates enhanced overlap between the Nb 4$d$ orbitals of neighboring clusters. This structural "unlocking" drives the expansion of the localization length ($\xi$) and the broadening of the bandwidth ($W$), ultimately overcoming the on-site Coulomb repulsion ($U$) to trigger the collapse of the Mott state. Consequently, the Jahn–Teller effect in GaNb$_4$Se$_8$ acts as the primary "orbital gate" controlling the transition from rigid, cluster-bound carrier confinement to a fully delocalized, correlated metallic regime.

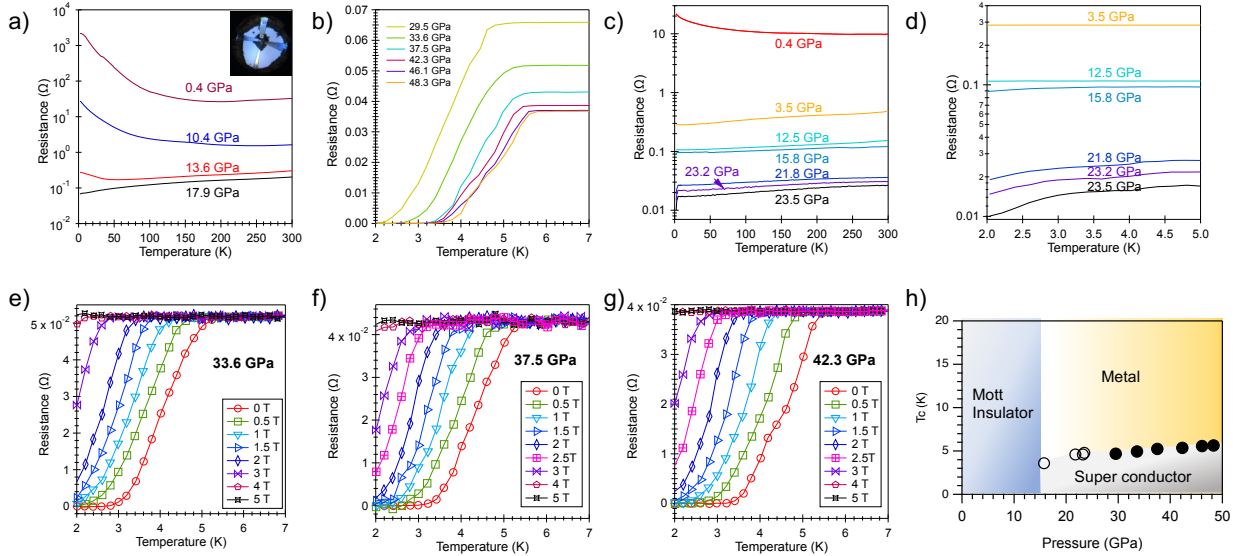

**Fig. 4.** Pressure-induced superconductivity in single-crystalline GaNb$_4$Se$_8$. (a) Electrical resistance as a function of temperature measured under compression. Inset: electrode configuration in the diamond anvil cell. (b) Expanded low-temperature resistance showing the emergence of superconductivity at high pressures. (c) Temperature-dependent resistance measured during decompression. (d) Enlarged view of panel (c) near the superconducting transition. (e–g) Magnetic-field dependence of the superconducting transition at selected pressures. (h) Pressure–temperature phase diagram summarizing the evolution from a Mott insulating state to a correlated metallic regime and superconductivity. Solid (open) symbols denote transition temperatures under compression (decompression).

Temperature-dependent resistivity measurements further clarify the pressure-driven electronic evolution (Fig. 4a). At low pressures, the resistance increases steeply upon cooling, characteristic of a Mott insulating state where carriers are strictly confined within Nb$_4$ clusters. With increasing pressure, the temperature dependence gradually weakens and transitions to a metallic regime between 10 and 14 GPa, confirming a pressure-induced insulator–metal transition. In contrast to the ES-VRH behavior observed at ambient pressure, the resistivity curves under compression systematically deviate from standard hopping models. This deviation suggests the emergence of metallic channels interwoven with thermally activated transport as the system approaches the correlated metallic regime.

In the intermediate crossover region (10–14 GPa), where traditional thermal activation and hopping models fail to provide a consistent fit, the resistivity curves are well-described by an empirical for $R(T)=R_0+Ae^{-BT}$, here, the parameter $B$ acts as a thermal sensitivity coefficient, reflecting the degree of carrier confinement within the Nb$_4$ network. As pressure increases, $B$ exhibits a systematic and dramatic reduction ( $0.0447$ and $0.00079\,\text{K}^{-1}$ at 10.4 and 13.6 GPa, respectively).

Physically, this attenuation of $B$ quantifies the progressive "softening" of the Mott insulating state. As the JT distortion is suppressed ($\sigma_{JT}\downarrow$) and the Nb $4d$ bandwidth ($W$) broadens, the energy barrier for inter-cluster hopping diminishes. Consequently, the resistance becomes increasingly insensitive to temperature fluctuations, signaling that the localization length is expanding beyond individual $Nb_4$ clusters. When $B$ approaches its minimum value at 13.6 GPa, the electronic wavefunctions are sufficiently delocalized to support a metallic-like framework, providing the necessary precursor for the bulk superconductivity observed at higher pressures.

At higher pressures, a sharp drop in resistance appears upon cooling, reaching a zero-resistance state between 29.5 and 48.3 GPa (Fig. 4b). This signals the emergence of bulk superconductivity, with the transition temperature $T_c$ increasing monotonically with pressure to exceed 5 K at 48.3 GPa. This evolution stems from the pressure-enhanced bandwidth ($W$) and inter-cluster coupling within the $Nb_4$ network, which facilitate long-range phase coherence. Notably, the establishment of a robust superconducting state near ~30 GPa is substantially higher than the ~13 GPa reported for polycrystalline samples [19]. However, since prior work did not observe zero resistance, we interpret their ~13 GPa threshold as the onset of superconducting fluctuations. In the present work, signatures of filamentary superconductivity are indeed observed at 17.9, 21.5, and 26.1 GPa (Supplementary Fig. S5). This distinction highlights the sensitivity of the superconducting percolation to microstructural order.

The intrinsic nature of the electronic reconstruction in $GaNb_4Se_8$ is further evidenced by the reversibility of the observed transitions. Upon decompression (Figs. 4c, d), the superconducting state gradually weakens and vanishes near ~12 GPa with the insulating behavior eventually reappearing at ambient pressure. This recovery indicates that the Mott-to-superconductor transition originates from an intrinsic electronic reconfiguration rather than irreversible structural damage, consistent with the restoration of the cubic symmetry upon pressure release. Notably, the transition pressures exhibited a pronounced hysteresis (~30 GPa during compression vs. ~12 GPa during decompression), a common feature in pressure-tuned correlated systems.

Magnetic-field-dependent resistivity measurements (Figs. 4e–g) confirm the high-pressure phase as a characteristic type-II superconductor. The superconducting transition is progressively suppressed with increasing magnetic field, and the upper critical field $\mu_0 H_{c2}(T)$ was analyzed using both the Ginzburg–Landau (GL) [25] and Werthamer–Helfand–Hohenberg (WHH) [26] models. Within the GL formalism, the temperature dependence is described by:

$$\mu_0 H_{c2}(T) = \mu_0 H_{c2}(0) \frac{1 - (T/T_c)^2}{1 + (T/T_c)^2}$$

while the WHH model in the dirty limit provides:

$$\mu_0 H_{c2}(0) = -0.693 T_c \frac{d(\mu_0 H_{c2})}{dT}\bigg|_{T=T_c}$$

The zero-temperature upper critical field $\mu_0 H_{c2}(0)$ was estimated between 4.26 T and 5.18 T across the 33.6–42.3 GPa range (Supplementary Fig. S6 for data fitting and Supplementary Table S1 for data summary). The resulting superconducting coherence length, $\xi(0)$, calculated via $\mu_0 H_{c2}(0) = \frac{\Phi_0}{2\pi \xi^2(0)}$, where $\Phi_0 = 2.07 \times 10^{-15}$ Wb, is approximately 8–9 nm. It is important to note that this $\xi(0)$ represents the spatial extent of the Cooper pairs in the delocalized metallic state, which is an order of magnitude larger than the electronic localization length ($\xi \approx 0.6$ nm) observed in the Mott insulating phase at low pressures.

The Pauli paramagnetic limit, $H_P \approx 1.84 T_c \approx 9$–10 T, remains significantly higher than the measured $\mu_0 H_{c2}(0)$, indicating that the upper critical field is primarily governed by orbital pair breaking. The slight increase in $\xi(0)$ with pressure may reflect the continued evolution of electronic correlations and the further broadening of the Nb $4d$ bandwidth ($W$).

The synthesized electronic phase diagram of GaNb4Se8 (Fig. 4h) reveals a sequential evolution: from a cluster Mott insulator governed by JT-induced confinement at ambient pressure, through a correlated metallic regime characterized by the suppression of $\sigma_{JT}$, and finally to a robust superconducting state. This pressure-temperature-tuned progression highlights the delicate interplay between local structural distortions and long-range electronic order in cluster-based Mott systems.

*Conclusion*—In summary, our study establishes the emergence of bulk superconductivity in single-crystalline $GaNb_4Se_8$ within a high-pressure monoclinic $C2$ framework. By utilizing high-quality single crystals, we identify the structural foundation for the transition from highly confined, cluster-bound carriers to a delocalized superconducting state.

Microscopically, we demonstrate that the pressure-induced suppression of JT distortions acts as a critical "orbital gate". This structural unlocking drives the expansion of the localization length $\xi$, facilitating the inter-cluster electronic overlap necessary for long-range phase coherence. Our results link the restoration of local symmetry to the collapse of the Mott state, identifying the "orbital gate" as a primary driver for overcoming electron correlations in cluster-based quantum materials

*Acknowledgments*—We acknowledge the support of GeoSoilEnviroCARS(Sector 13), which is supported by the National Science Foundation−Earth Sciences (EAR-1634415). This research used resources of the Advanced Photon Source, a U.S. Department of Energy (DOE) Office of Science User Facility operated for the DOE Office of Science by Argonne National Laboratory under contract No. DE-AC02-06CH11357. Use of the COMPRES-GSECARS gas-loading system and 13-BM-C was supported by COMPRES under NSF Cooperative Agreement No. EAR-1661511 and by GSECARS through NSF Grant No. EAR-1634415 and DOE Grant No. DE-FG02-94ER14466. L.W. acknowledges the support by the Natural Science Foundation of China (Grants No. 52288102 and No. 52090020) and the support from the S&T Program of Hebei (225A1102D). L.P. and V.T acknowledge the support of the Deutsche Forschungsgemeinschaft (DFG) through Transregional Research Collaboration TRR 360 (project No. 492547816) and by the project ANCD (cod 011201 Moldova). D.Z. acknowledges the support of NSF EAR-2246686. Y.W. is grateful to the Faculty Research Grant from Oakland University for supporting this research.

Supplementary Information

# From Wavefunction Collapse to Superconductivity: Evolution of the Electronic State in Compressed GaNb$_4$Se$_8$

Yuejian Wang[1,*], Zhongyan Wu[2], K C Bhupendra[3], Dongzhou Zhang[4], Lin Wang[2], Sanjay V. Khare[3], Lilian Prodan[5,6], Vladimir Tsurkan[5,6]

[1]Department of Physics, Oakland University, Rochester, MI 48309, USA
[2]Center for High-Pressure Science (CHiPS), State Key Laboratory of Metastable Materials Science and Technology, Yanshan University, Qinhuangdao, Hebei 066004, China
[3]Department of Physics and Astronomy, and Wright Center for Photovoltaics Innovation and Commercialization (PVIC), University of Toledo, Toledo, Ohio 43606, USA
[4]GeoSoilEnviroCARS, University of Chicago, Argonne, IL 60439, USA
[5]Experimental Physics V, Center for Electronic Correlations and Magnetism, University of Augsburg, Augsburg 86135, Germany
[6]Institute of Applied Physics, Moldova State University, MD-2028 Chisinau, Republic of Moldova

*Email: Ywang235@oakland.edu

**Table of Contents**

**Supplementary Note 1: Synthesis and Characterization**

Polycrystalline GaNb$_4$Se$_8$ was synthesized by a conventional solid-state reaction using high-purity elements: gallium pellets (99.9999%, Alfa Aesar), niobium powder (325 mesh, 99.8%, Alfa Aesar), and selenium granules (2–4 mm, 99.999%, ChemPUR). Stoichiometric mixtures of these elements were pressed into pellets, sealed under vacuum (~10$^{-3}$ mbar) in double quartz ampoules, with the inner ampoule carbon-coated to prevent reactions between the sample and quartz. The sealed ampoules were heated in a muffle furnace according to the following sequence: ramped at 100 °C/h to 600 °C and held for 10 h; then ramped at 20 °C/h to 700 °C (dwelling for 20 h), 20 °C/h to 800 °C (25 h dwell), and finally 10 °C/h to 900 °C with a 68 h dwell before cooling to room temperature at 100 °C/h.

After the first sintering cycle, phase analysis was performed using a STOE Stadi P diffractometer with Cu Kα radiation. The diffraction pattern revealed a predominant GaNb$_4$Se$_8$ phase with minor traces (<10%) of hexagonal NbSe$_2$ impurity (Supplementary Fig. S1b) [1]. The product was subsequently reground, re-pelletized, and annealed again at 900 °C for an additional 72 h. The second cycle yielded a phase-pure GaNb$_4$Se$_8$ (Supplementary Figs. S1b, c).

Single crystals were grown by the chemical vapor transport (CVT) method using the as-synthesized polycrystalline material as a precursor. Approximately 2 g of GaNb$_4$Se$_8$, together with 0.2 g of iodine (as a transport agent) and 100 mg of selenium granules (to maintain stoichiometry), were sealed in an evacuated quartz ampoule (18 mm × 130 mm). Crystal growth was conducted in a temperature gradient between 1000 °C (hot zone) and 960 °C (cold zone) for ~12 weeks, which was necessary for complete mass transport. The process yielded high-quality single crystals of well-defined cubic morphology with shiny (111) and (001) facets (Supplementary Fig. S1a), as well as hexagonal plate-like crystals formed in the cold zone.

The phase purity and elemental composition of the single crystals were examined using a ZEISS Crossbeam 550 scanning electron microscope (SEM) equipped with energy-dispersive X-ray spectroscopy (EDS). The cubic crystals exhibited homogeneous single-phase composition (Supplementary Fig. S1d) with an average stoichiometry of Ga$_{1.05(2)}$Nb$_4$Se$_{8.01(4)}$, assuming full Nb-site occupancy. The hexagonal plates were identified as NbSe$_2$ slightly doped with Ga (~0.07%).

Magnetic susceptibility (χ) was measured using a Quantum Design MPMS 3 SQUID magnetometer on a larger single crystal from the same batch, with a magnetic field of 7 T applied along the [001] direction (Supplementary Fig. S1e), in excellent agreement with previous studies [2-4].

**Supplementary Note 2: High-pressure Powder X-ray Diffraction**

High-pressure X-ray diffraction experiments were performed using a symmetric diamond anvil cell (DAC) equipped with 300 μm culets. Powder samples were loaded into a hole (serving as the sample chamber) with a diameter of 130 μm and a depth of ~45 μm, drilled at the center of a rhenium gasket pre-indented by diamond anvils. A ruby sphere (<15 μm in diameter) was placed near the sampling material in the chamber for pressure determination [5].

High-pressure powder X-ray diffraction (PXRD) experiments were conducted at 13-BM-C beamline of GSECARS [6]. The beamline was calibrated using a LaB$_6$ standard to determine the sample-to-detector distance and detector tilt angle.

After the beamline calibration, an initial X-ray diffraction pattern was collected under ambient conditions to evaluate the crystal structure and sample quality. To minimize nonhydrostatic effects during measurements, helium gas was loaded using the gas-loading system at GSECARS (Argonne National Laboratory), serving as the pressure-transmitting medium to ensure quasi-hydrostatic conditions [7]. During the subsequent sealing of the chamber with helium gas, the pressure was maintained as low as possible (below 1 GPa) to enable data collection from the lowest attainable pressure point.

Pressure was increased using a membrane-driven system which significantly improved the efficiency and accuracy of the measurement. Upon reaching each target pressure, the DAC was stabilized for approximately 10 minutes to allow the sample to reach a quasi-equilibrium state. Immediately after the 10-minute relaxation period, the sample pressure was measured and the the X-ray diffraction pattern was recorded. The pressure was then re-measured immediately after data acquisition, and the average of the two readings (before and after pattern collection) was taken as the effective sample pressure.

The collected diffraction rings were converted into conventional one-dimensional patterns (intensity vs. 2θ) using Dioptas, allowing the evolution of diffraction features to be monitored in real time [8]. Subsequently, the X-ray diffraction patterns were refined using GSAS-II to extract the lattice parameters and unite-cell volume at each pressure point [9]. Overall, two key phenomena were identified within the sample: the onset of a phase transition at approximately 20 GPa, which completed near 30 GPa, and the exceptional incompressibility inferred from equation-of-state analysis.

**Supplementary Note 3**: **High-pressure Single-Crystal X-ray Diffraction**

To identify the crystal structure of the high-pressure phase, single-crystal X-ray diffraction (SCXRD) measurements were conducted at beamline 13-BM-C. A SS-80 diamond anvil cell (purchased from DACTools) equipped with two 300 µm culets was used as the pressure cell. A 200 µm-thick rhenium gasket was pre-indented to ~45 µm, and a 200 µm-diameter hole was drilled at the center of the indentation. A small single crystal (~30 µm) together with a ruby sphere (pressure calibrant) were loaded into the chamber together. Neon was loaded as the pressure-transmitting medium using the GSECARS gas-loading system [7].

The X-ray beam was monochromated to λ = 0.4306 Å with a 1 eV bandwidth using a Si(311) monochromator, and focused to a 9 µm × 11 µm (FWHM) spot by a Kirkpatrick–Baez mirror system. Diffracted intensities were recorded with a Pilatus3 1M detector (Dectris) positioned ~200 mm from the DAC. A NIST SRM 660a $LaB_6$ standard was used to calibrate the sample–to–detector distance and detector tilt. Prior to data collection, the sample was centered at the rotation axis of the diffractometer. Compression was controlled by a membrane-driven system. Diffraction patterns were collected over a 70° rotation range with 0.5° steps and 1 s exposure time per frame.

The diffraction images were integrated and processed using APEX 5 (Bruker). Intensities were scaled according to the diffracted volumes at different rotation angles. The corrected reflection intensities were used for crystal-structure refinement with Olex2, which incorporates SHELXL [10, 11].

**Supplementary Note 4: Electrical Transport Measurements**

Electrical transport measurements under magnetic fields were performed using a cryogen-free C-Mag superconducting magnet system (Cryomagnetics, Inc.). For combined low-temperature and high-pressure measurements, a DAC equipped with 200 μm culets was employed. A Be–Cu gasket was pre-indented, and a ~ 200 μm-diameter hole was drilled at the center of the indentation. The hole was filled with a cubic boron nitride (cBN)/epoxy insulating mixture and further compressed. A ~120 μm-diameter hole was subsequently drilled to serve as the sample chamber. A single crystal of GaNb$_4$Se$_8$ (~70 μm in diameter and ~5 μm thick) was loaded together with NaCl as the pressure-transmitting medium and four platinum electrodes (~2 μm thick). The electrodes were connected to Cu leads using Ag paste. Electrical resistance was measured in a standard four-probe configuration with an excitation current of 100 μA and a temperature sweep rate of 2 K/min.

Pressure was calibrated at room temperature using ruby fluorescence before cooling and after warming in each cycle, with a ~5 μm ruby sphere positioned near the sample. Owing to thermal expansion of the diamond-anvil cell, cooling typically induced a pressure increase of ~1 GPa; the pressure measured after warming was therefore used as the effective value. Representative R(T) curves (Supplementary Fig. S7) show mild cooling–warming hysteresis. Data acquired during controlled warming (2 K/min) are smoother and less sensitive to thermal pressure drift, suggesting that pressure variations do not significantly affect the determination of superconducting transition temperatures.

**Supplementary Note 5: Technical Details of Transport Analysis in GaNb$_4$Se$_8$**

**1. The Shklovskii-Efros (SE) Variable-Range Hopping (VRH) Model**

The temperature-dependent resistivity $\rho(T)$ is analyzed using the SE-VRH model [14, 15], which accounts for the Coulomb gap in the density of states near the Fermi level:

$$\rho(T) = \rho_0 \cdot e^{(\frac{T_0}{T})^{1/2}} \qquad (1)$$

Where $T_0$ is the characteristic ES temperature and $\rho_0$ is the pre-exponential factor. To extract these parameters, Eq. (1) is linearized by taking the natural logarithm:

$$\ln(\rho) = \ln(\rho_0) + \sqrt{T_0} \cdot T^{-0.5} \qquad (2)$$

The characteristic temperature $T_0$ is determined from the square of the slope (m) of the $\ln(\rho)$ vs. $T^{-0.5}$ plot: $T_0 = m^2$.

**2. Calculation of the Localization Length ($\xi$)**

The localization length $\xi$, representing the spatial extent of the electronic wavefunction, is derived from $T_0$ using the following relation:

$$\xi = \frac{\beta \cdot e^2}{4\pi\epsilon_0 \varepsilon_r k_B T_0} \qquad (3)$$

In our calculations, we utilize the following standard constants:
- Numerical coefficient **$\beta$=2.8**.
- Relative dielectric constant **$\varepsilon_r \approx$10 [16].**

- Physical constants:

    $e=1.602\times10^{-19}$ C, $k_B=1.381\times10^{-23}$ J/K, and $\epsilon_0=8.854\times10^{-12}$ F/m.

For practical computation in units of Ångströms (Å), Eq. (3) simplifies to:
$$\xi(\text{Å}) \approx \frac{467{,}600}{\varepsilon \cdot T_0(K)} \quad (4)$$

### 3. Calculation of the Hopping Energy ($W_{ES}$)

The average hopping energy $W_{ES}$, representing the energy barrier for a carrier jump at a specific temperature $T$, is calculated as:
$$W_{ES}(T) = 0.5 \cdot k_B \cdot \sqrt{T_0 \cdot T} \quad (5)$$
Using our fitted values, we obtain $W_{ES} \approx 38.2\ meV$ at 260 K (High-T regime) and $W_{ES} \approx 32.7\ meV$ meV at 75 K (Low-T regime).

### 4. Comparison with Lattice Geometry

The inter-cluster distance $d_{inter}$ between adjacent $Nb_4$ tetrahedra is calculated from the cubic lattice constant $a=10.42$ Å (at 300 K) as measured from the x-ray diffraction:
$$d_{inter} = \frac{a}{\sqrt{2}} \approx 7.4\text{Å} \quad (6)$$
Accounting for structural contraction at low temperatures, $d_{inter}$ is estimated to be $\approx 6.7$ Å, which serves as the geometric baseline to confirm wavefunction confinement ($\xi \approx 6.1\text{Å}$).

**Supplementary Note 6: Computational Methods**

Density functional theory (DFT) calculations were performed using the Vienna Ab Initio Simulation Package (VASP) [17-19]. The Perdew-Burke-Ernzerhof (PBE) [20] Generalized Gradient Approximation (GGA) exchange-correlation functional was employed within the Projector-Augmented-Wave (PAW) framework [18, 21, 22]. Nb_sv, Ga_d, and Se PAW pseudopotentials were employed with a plane-wave cutoff energy of 520 eV. Brillouin-zone sampling was performed using a 5 × 5 × 5 k-point mesh, verified to be converged. Gaussian smearing with a width of 0.05 eV was applied to set the partial occupancies [23]. Structures were fully relaxed until the residual forces were below 0.01 eV Å$^{-1}$, with an electronic self-consistency criterion of $10^{-6}$ eV.

The 0 GPa structure was obtained from the Materials Project database [24]. The high-pressure structure was obtained by relaxing the unit cell at the target pressure starting from the 0 GPa cubic phase. Optimized lattice parameters at 0 and 15 GPa are summarized in Supplementary Table S2. To account for correlation effects in Nb 4d states, DFT+U calculations were performed using an effective Hubbard parameter U = 6 eV, consistent with previous studies [25, 26]. Hybrid-functional (HSE06) calculations were additionally carried out at ambient pressure to obtain an improved estimate of the electronic band gap. Due to the large unit cell (52 atoms), pressure-dependent electronic structure calculations were restricted to the GGA+U level.

Chemical bonding analysis was performed using the crystal orbital Hamilton population (COHP) method as implemented in the Local Orbital Basis Suite Towards Electronic-Structure Reconstruction (LOBSTER) code [27-29]. Projected COHP (pCOHP) analysis was carried out using a distance window of 2.5–4.5 Å to capture first-neighbor Nb–Se and Ga–Se interactions. Positive and negative −pCOHP values correspond to bonding and antibonding states, respectively, with the Fermi level set to 0 eV. Phonon calculations were performed using Phonopy [30] to assess the thermodynamic stability of candidate crystal structures under pressure (see Supplementary Table S3).

**Supplementary Table S1.** Critical superconducting temperatures under applied magnetic fields and estimated upper critical fields.

|  | 33.6 GPa | | 37.5 GPa | | 42.3 GPa | |
|---|---|---|---|---|---|---|
|  | onset | 50% drop | onset | 50% drop | onset | 50% drop |
| Field (T) | $T$ (K) | | | | | |
| 0 | 4.94 | 4.05 | 5.24 | 4.33 | 5.37 | 4.64 |
| 0.5 | 4.5 | 3.62 | 4.713 | 3.91 | 4.822 | 4.11 |
| 1 | 4.03 | 3.29 | 4.296 | 3.5 | 4.254 | 3.66 |
| 1.5 | 3.65 | 2.97 | 3.833 | 3.2 | 3.791 | 3.26 |
| 2 | 3.28 | 2.64 | 3.474 | 2.85 | 3.349 | 2.81 |
| 2.5 | - | - | 3.08 | 2.49 | 2.984 | 2.42 |
| 3 | 2.57 | 1.98 | 2.72 | 2.04 | 2.57 | 1.99 |
|  | $H_{c2}$ (0) (T) | | | | | |
| GL | 5.18 | 4.91 | 5.14 | 4.88 | 4.67 | 4.35 |
| WHH |  | 4.83 |  | 4.86 |  | 4.26 |

**Supplementary Table S2**. Lattice parameters of cubic GaNb$_4$Se$_8$ obtained from GGA calculations.

| Pressure (GPa) | Lattice constant (a) Å |
|---|---|
| 0 | 10.36, 10.32[a], 10.42[b] |
| 15 | 10.17 |

[a] Experimental Ref. [31]
[b] Experimental Ref. [26]

**Supplementary Table S3.** Helmholtz free energy (F) of cubic and orthorhombic $GaNb_4Se_8$ structures as a function of pressure at 300 K.

| Pressure (GPa) | Cubic (kJ/mol) | Orthorhombic (kJ/mol) |
|---|---|---|
| 00 | -80.62 | -61.97 |
| 05 | -59.13 | -51.27 |
| 10 | -54.70 | -39.53 |
| 15 | -50.49 | -27.91 |
| 20 | -62.66 | -25.12 |
| 25 | -49.42 | -22.24 |
| 30 | -37.96 | -21.19 |
| 35 | -28.24 | -16.95 |
| 40 | -19.91 | -13.02 |
| 45 | -12.54 | -9.60 |

**Figures:**

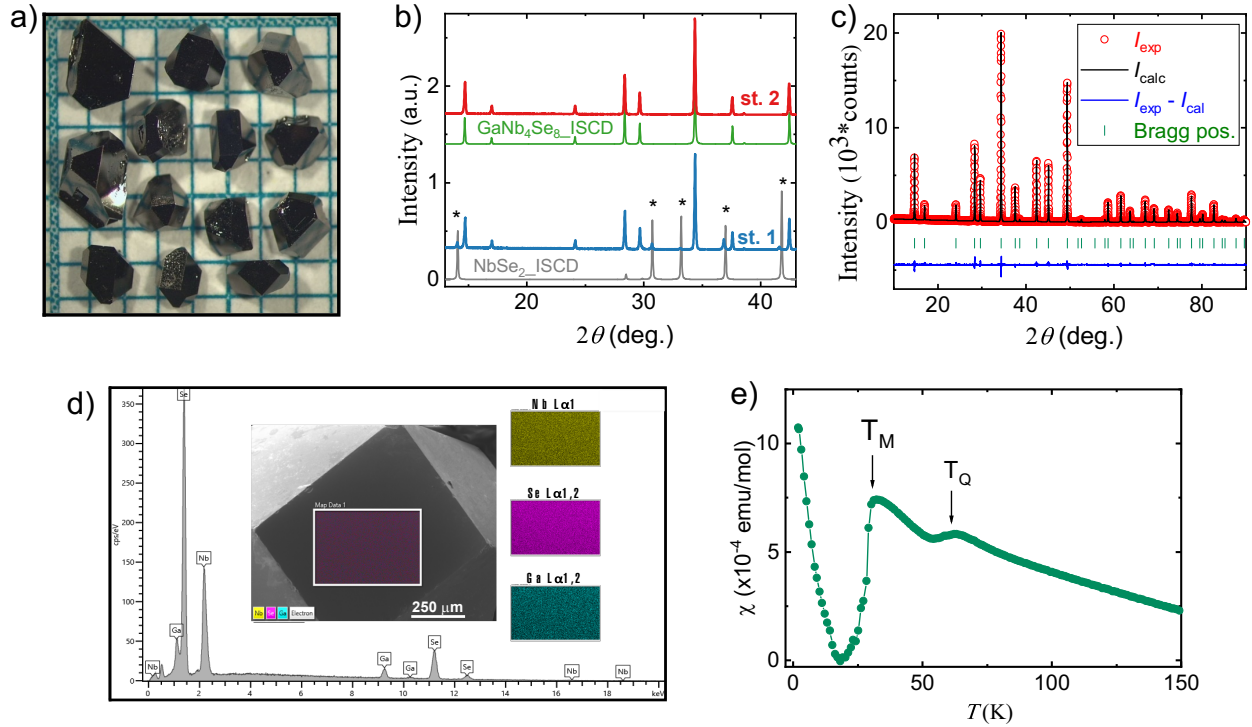

**Supplementary Figure S1. Crystal growth and characterization.** a) Optical image of as-grown GaNb$_4$Se$_8$ crystals showing well-defined (111) triangular and (001) rectangular facets. b) X-ray diffraction patterns of polycrystalline samples after the first (st.1) and second (st.2) sintering cycles, compared with reference patterns of NbSe$_2$ and GaNb$_4$Se$_8$ from the Inorganic Crystal Structure Database (ICSD, FIZ Karlsruhe). Asterisks mark diffraction peaks from the NbSe$_2$ impurity phase. c) Rietveld refinement of the st.2 sample confirming the formation of a single-phase cubic GaNb$_4$Se$_8$ (space group F$\bar{4}$3m). d) Energy-dispersive X-ray (EDS) spectrum acquired on the (001) surface of an as-grown GaNb$_4$Se$_8$ crystal. Insets show scanning electron microscopy (SEM) elemental maps illustrating the homogeneous chemical composition. e) Temperature dependence of magnetic susceptibility, showing distinct λ-type anomalies at the spin-singlet transition T$_M$ and the quadrupolar transition T$_Q$.

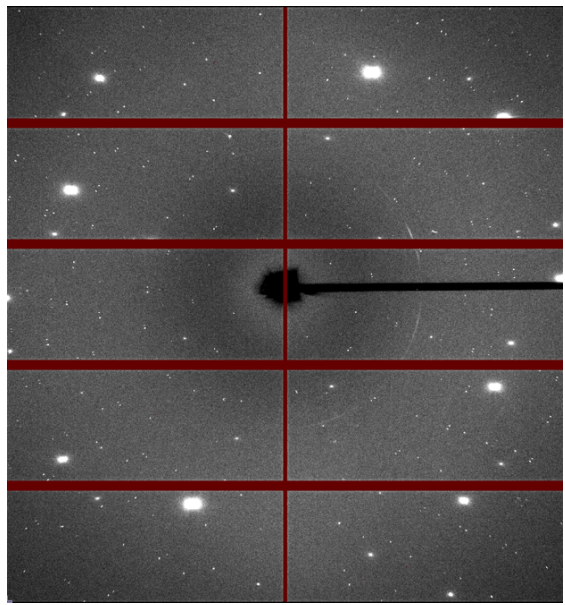

**Supplementary Figure S2.** Single-crystal diffraction pattern of GaNb$_4$Se$_8$ at ambient pressure. Red regions correspond to masked detector shadows.

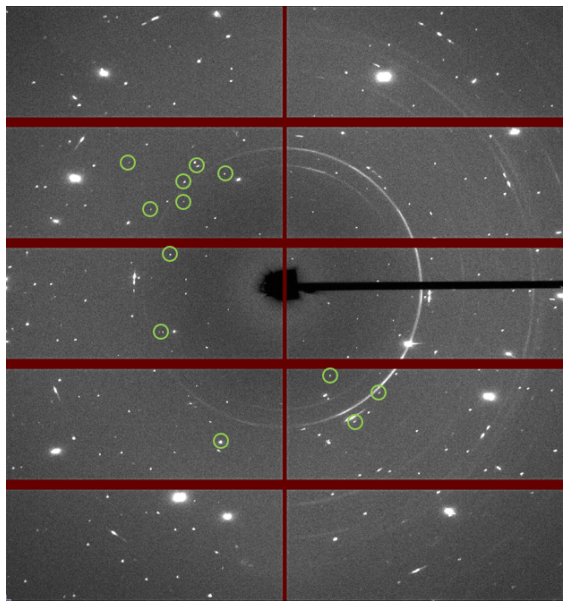

**Supplementary Figure S3.** Single-crystal diffraction pattern of GaNb$_4$Se$_8$ at 21.0 GPa. Green circles highlight the diffraction spots emerging from the new phase.

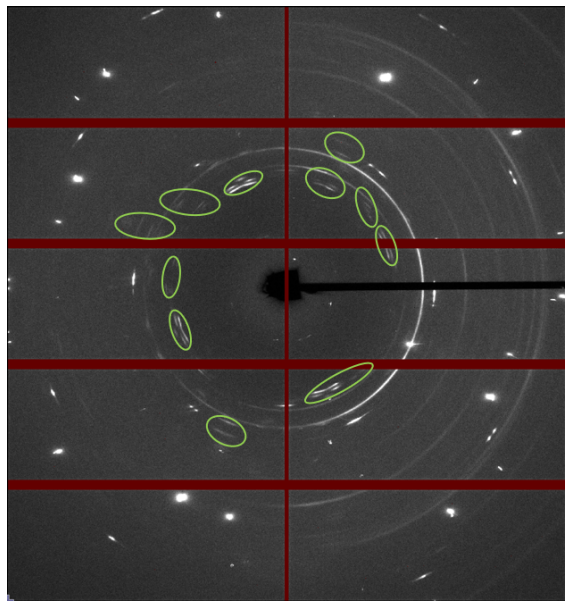

**Supplementary Figure S4.** Diffraction pattern of GaNb$_4$Se$_8$ at 28.2 GPa. The crystal is crushed into powder under this pressure, and the green ellipses mark the diffraction reflections of the high-pressure phase.

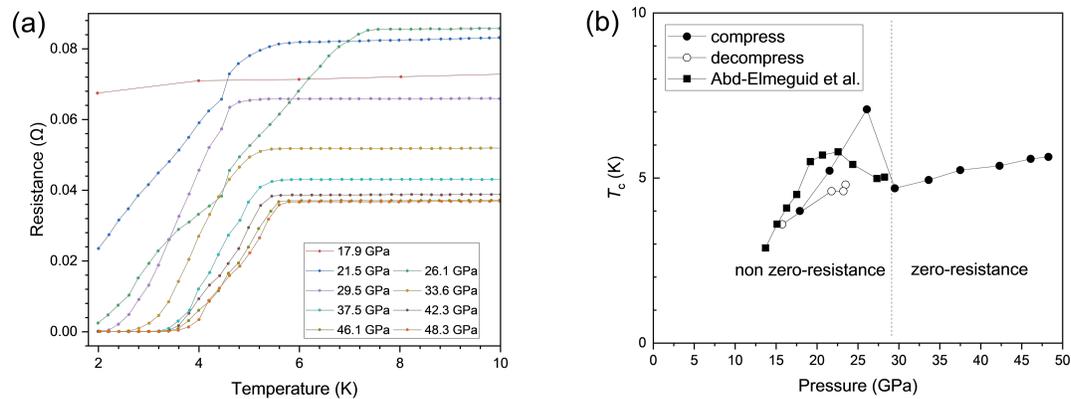

**Supplementary Figure S5.** (a) Temperature-dependent resistance of GaNb$_4$Se$_8$ measured between 2 and 10 K at pressures from 17.9 to 48.3 GPa. (b) Pressure dependence of the superconducting critical temperature ($T_c$). Temperature-dependent resistance at intermediate pressures, 17–26 GPa, showing filamentary superconductivity.

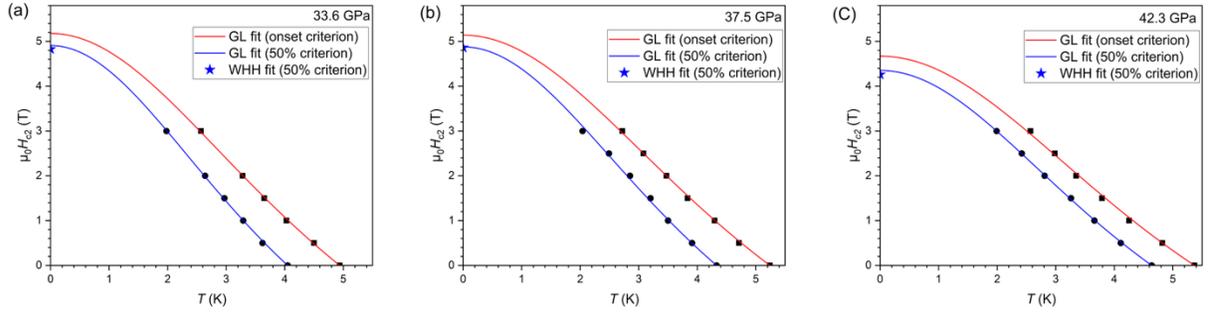

**Supplementary Figure S6.** Estimation of the upper critical magnetic field ($\mu_0 H_{c2}$) at 33.6 (a), 37.5 (b), and 42.3 GPa (c). The critical superconducting transition temperatures ($T_c$) at different magnetic fields were fitted using the Ginzburg–Landau (GL) and Werthamer–Helfand–Hohenberg (WHH) models. Square and circle symbols denote $T_c$ values determined from the onset of the resistance drop and the 50% resistance-drop criterion, respectively.

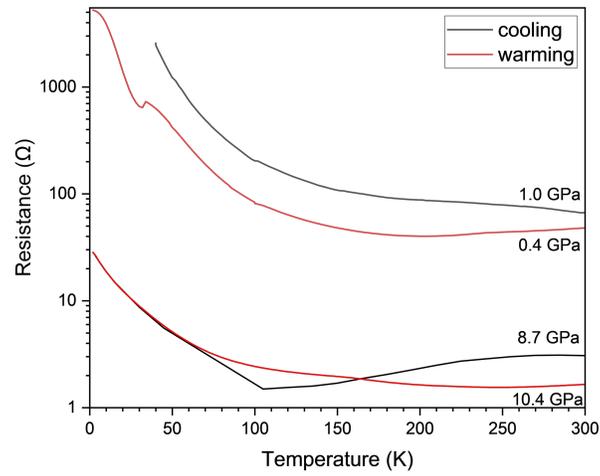

**Supplementary Figure S7.** Comparison of electrical transport data collected during cooling and warming cycles.